\documentclass[doublecol]{epl2}

\title{Minimal conductivity in graphene: interaction corrections and ultraviolet anomaly}

\author{E.G.~Mishchenko}

\institute{
  Department of Physics, University of Utah, Salt Lake
City, Utah 84112, USA\\
} 

\abstract{Conductivity of a disorder-free intrinsic graphene is
studied to the first order in the long-range Coulomb interaction and
is found to be $\sigma=\sigma_0(1+0.01 g)$, where $g$ is the
dimensionless (``fine structure'') coupling constant. The
calculations are performed using three different methods: i)
electron polarization function, ii) Kubo formula for the
conductivity, iii) quantum transport equation. Surprisingly, these
methods yield {\it different} results {\it unless} a proper
ultraviolet cut-off procedure is implemented, which requires that
the interaction potential in the effective Dirac Hamiltonian is
cut-off at small distances (large momenta).}

\begin{document}

\maketitle

\section{Introduction} Low-frequency optical conductivity of undoped
(intrinsic)  graphene free of disorder is known to have a universal
value of $\sigma_0 = e^2/4\hbar$ \cite{fradkin,
lee,ludwig,shon,gusynin,kats,jakub,cserti,nomura,falk,peres,ziegler,akira}.
Experimental measurements \cite{Geim,Nov}, which yielded a value
somewhat bigger than the theoretical predictions, motivated the
studies of the possible role played by electron-electron
interactions. The findings of Ref.~\cite{2} that the combined effect
of self energy (velocity renormalization) and vertex corrections
leads to a suppression of the optical conductivity at low
frequencies have been questioned in Refs.~\cite{sheehy,herbut} on
the basis of scaling arguments. The latter indicate that the large
logarithmic (momentum cut-off dependent) terms in the self-energy
and vertex corrections cancel each other. We note that Ref.~\cite{2}
and Refs.~\cite{sheehy,herbut} agree on this cancellation in the
lowest order in electron-electron interaction but differ on whether
the higher order terms feature similar cancellation. It appears that
the analysis of Ref.~\cite{2}, though valid in the first order,
fails for higher orders, and that the conclusion of the suppression
of the conductivity at low frequencies is not valid.

The theory presented in Refs.~\cite{sheehy,herbut} implies that the
low-frequency dependence is properly described by the lowest order
correction. Indeed, to the first order in interaction the
conductivity is expected to yield, $\sigma/\sigma_0 = 1+Cg$, where
$C$ is some constant, $g=e^2/\kappa v$ is the interaction strength;
$\kappa$ is the dielectric constant of a substrate and $v$ is the
electron velocity in graphene. Renormalization group approach for 2D
Dirac fermions predicts that the interaction strength $g$ is a
running coupling constant that depends on frequency $g \to
\widetilde g(\omega)$ \cite{Gonzales,Gonzales1}. At low frequencies
$\widetilde{g}(\omega)$ flows to zero, so that higher order
corrections to the electron velocity become progressively negligible
and it is sufficient to consider only the first order
renormalization of velocity (electric charge is not renormalized):
$\widetilde g(\omega)=g/[1+ \frac{g}{4}\ln{({\cal K}v/\omega})]$,
where ${\cal K}$ is the momentum cut-off. Combining these
expressions gives,
\begin{equation}
\label{form} \sigma/\sigma_0=  1+\frac{Cg}{1+ \frac{g}{4}\ln{({\cal
K}v/\omega})},
\end{equation}
with the low-frequency behavior of the conductivity being determined
by the constant $C$ alone. Calculation of this constant, therefore,
becomes an important task. While Ref.~\cite{sheehy} did not
calculate $C$, Ref.~\cite{herbut} provided the following value
\begin{equation}
\label{Cher} C=\frac{25-6\pi}{12}\approx 0.51~.
\end{equation}
This result predicts quite a considerable variation of $\sigma$ with the
frequency for typical values of the bare graphene interaction constant $g$
(which can exceed 1).

In the present Letter we test the above prediction (\ref{Cher}) by
performing a perturbative calculation of the minimal conductivity to
the first order in electron-electron interaction using three
different methods, based on, a) electron polarization operator, b)
Kubo formula for the conductivity, c)  kinetic equation. We point
out that crucial anomaly, which does not appear in a non-interacting
case, occurs for the interaction correction. Three above mentioned
methods would give essentially {\it different} values for the
constant ${\cal C}$ unless some appropriate high-momentum cut-off
procedure is implemented. We argue that expression (\ref{Cher})
overestimates the interaction correction by almost two orders of
magnitude and show that the numerical value of $C$ is
\begin{equation}
\label{result} C=  \frac{19-6\pi}{12}\approx 0.01~.
\end{equation}
We will now proceed to demonstrate that the difference between
Eqs.~(\ref{Cher}) and (\ref{result}) originate from handling of
singular integrals at large electron momenta.

The first method to be presented is based on the calculation of
electron polarization operator and has an advantage of {\it  being
free} from any such singular integrals.
\begin{figure}
\onefigure{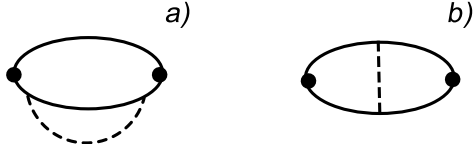} \caption{Self-energy, a), and vertex
correction, b), to the conductivity $\sigma(\omega)$ and
polarization operator $\Pi(\omega,q)$
 in the first order in
electron-electron interaction (dashed line). The vertex (black dot)
 is equal to $1$ in the case of the polarization operator and
to $ev\hat {\bm \sigma}$ in case of the conductivity. The two
quantities are related to each other by the particle conservation
condition, Eq.~(\ref{conduc}).} \label{fig.1}
\end{figure}

\section{Polarization operator} Single intrinsic 2D graphene layer is
described by the chiral Hamiltonian,
\begin{equation}
\label{ham} H=v\sum_{i \bf p} ~ \hat c^{i\dagger}_{{\bf p}} \hat
{\bm \sigma}\cdot {\bf p} ~\hat c^{i}_{\bf p}+\frac{1}{2}\sum_{ij
\bf pkq} \hat c^{i\dagger}_{ {\bf p-q}}\hat c^{j\dagger}_{{\bf k+q}}
V_{\bf q} \hat c^j_{\bf k} \hat c^i_{{\bf p}},
\end{equation}
where ``hats'' denote operators in a pseudo-spin space ($\hat{\bm
\sigma}$ represents the usual set of Pauli matrices), the sum over
Latin indices is taken over two nodal points and two (true) spin
directions. The interaction potential is $V_{\bf q}=2\pi e^2/\kappa
q$; we also denote, $\sum_{\bf p} \equiv \int {d^2p}/{(2\pi)^2}$,
and set $\hbar=1$.

First-order interaction corrections to the conductivity are given by
the two diagrams  shown in Fig.~1, with the vertices denoting the
operators of electric current, $ev\hat {\bm \sigma}$. Another
possible method
 to derive the homogeneous optical conductivity is to calculate the corresponding diagrams for the
electron {\it polarization operator} $\Pi(\omega,q)$ and then
utilize the particle conservation condition,

\begin{eqnarray} \label{conduc} \sigma (\omega)
=\lim_{q \to 0} \frac{ie^2\omega}{q^2} \Pi (\omega,q).
\end{eqnarray}
The calculation of the polarization operator to the first order in
$g$ requires two diagrams \cite{note},
\begin{eqnarray}
\label{pi_start}
 \Pi (\omega, q)=4 Tr \sum_{\bf p p'}
\int\frac{d\epsilon d\epsilon'}{(2\pi)^2} ~V_{\bf p-p'}\Bigl[ 2 \hat
G_{\epsilon \bf p}
\hat G_{\epsilon' \bf p'}\hat  G_{\epsilon \bf p}  \nonumber\\
\times \hat G_{\epsilon+\omega, \bf p+q}+\hat  G_{\epsilon \bf p}
\hat G_{\epsilon' \bf p'}\hat  G_{\epsilon'+\omega, \bf p'+q} \hat
G_{\epsilon+\omega, \bf p+q} \Bigr],
\end{eqnarray}
where the electron Green's function in the subband representation is
\begin{equation}
 \label{subband}
 \hat G_{\epsilon {\bf p}}=\frac{1}{2} \sum_{\beta=\pm 1} \frac{1+\beta \hat \sigma_{\bf
 p}}{\epsilon-\beta(
 vp-i\eta)},
 \end{equation}
with $\hat\sigma_{\bf p}=   \hat{\bm \sigma}\cdot {\bf n}_{\bf p}$
being the
 projection of the pseudo-spin operator onto the direction of the
 electron momentum ${\bf n}_{\bf p}={\bf p}/p$. Factor 4 in
 Eq.~(\ref{pi_start}) accounts for the (true) spin and valley degeneracy.
 Taking energy integrals and performing pseudospin trace operation
 we obtain for the first term in Eq.~(\ref{pi_start}),
\begin{equation}
\label{pia} \Pi_{a} (\omega, q)=2\sum_{\bf p p'\beta}~V_{\bf p-p'}
\cos{\theta_{\bf p, p'}} \frac{1-\cos{\theta_{\bf p,
p+q}}}{(\omega+2\beta vp-i\beta \eta)^2}.
\end{equation}
In the expression (\ref{pia}) we kept only terms that lead to the
lowest order contribution in the external momentum, $\Pi_{a}
(\omega, q) \propto q^2$, which are necessary for the calculation of
the homogeneous conductivity. Using Eq.~(\ref{conduc}) we obtain the
corresponding contribution,
\begin{eqnarray}
\label{a} \sigma_{a}(\omega)=ie^2\omega \int \sum_{\bf p p'}~V_{\bf
p-p'} \cos{\theta}_{\bf p, p'} \frac{\omega^2+4v^2p^2}{p^2
(\omega^2-4v^2p^2)^2},
\end{eqnarray}
where the frequency is presumed to have a positive infinitesimal
imaginary part. The second term in Eq.~(\ref{pi_start}) is evaluated
similarly,
\begin{eqnarray}
\Pi_{b} (\omega, q)=-2 \sum_{\bf p p'}\left( \frac{ \omega^2 ({\bf
n}_{\bf p} -{\bf n}_{\bf p+q})\cdot ({\bf n}_{\bf p'} -{\bf
n}_{\bf p'+q})}{(\omega^2-4v^2p^2)(\omega^2-4v^2p'^2)} \right.\nonumber\\
\left. +\frac{4v^2pp' \sin{\theta}_{\bf p, p+q}\sin{\theta}_{\bf p',
p'+q}}{(\omega^2-4v^2p^2)(\omega^2-4v^2p'^2)} \right) V_{\bf p-p'}.
\end{eqnarray}
Expanding to the quadratic order in $q$ we obtain the vertex
correction,
\begin{eqnarray}
\label{b} \sigma_{b}(\omega)=-ie^2 \omega \sum_{\bf p p'} V_{\bf
p-p'} \frac{\frac{\omega^2}{pp'} \cos^2{\theta}_{\bf p, p'} +4v^2
\cos{\theta}_{\bf p, p'}}{(\omega^2-4v^2p^2)(\omega^2-4v^2p'^2)}.
\end{eqnarray}

To the first order in interaction the conductivity is given by the
sum $\sigma=\sigma_0+\sigma_{a}+\sigma_{b}$. The second term here,
given by  Eq.~(\ref{a}), contains a strong divergence at
$p=\omega/2v$. This divergence, however, is simply a consequence of
the renormalization of the electron velocity by electron-electron
interactions. To make the integrals regular we note that {\it both}
the zeroth-order term \cite{note} and $\sigma_{a}$ can be written as
\begin{equation}
\label{two} \sigma_0+\sigma_{a}=2ie^2\omega \sum_{\bf p}\frac{v_p}
{p~(\omega^2-4v_p^2p^2)},
\end{equation}
with $v_p=v+\frac{1}{2p}\sum_{\bf p'}V_{\bf p-p'}\cos{\theta}_{\bf
p, p'} =v[1+\frac{g}{4}\ln{({\cal K}/p})]$ being the  renormalized
velocity (where ${\cal K}$ is the upper momentum cut-off). Indeed,
expanding the integrand to the first order in the interaction one
recovers Eq.~(\ref{a}). Note that the value of $v_p$ coincides with
the electron velocity found from the perturbation expansion for the
electron Green's function \cite{Gonzales}. The integral in
Eq.~(\ref{two}) is regular. Calculating the real part of
Eq.~(\ref{two}) we obtain
\begin{equation}
\label{self} \sigma_0+\sigma_a'=\sigma_0\left(1+\frac{g}{4}\right).
\end{equation}
Note that the interaction correction in Eq.~(\ref{self}) is due to
the {\it curvature} of electron spectrum.

 Calculation of the real
part of Eq.~(\ref{b}) can be reduced to the following dimensionless
integral ($x=2vp/\omega$),
\begin{eqnarray}
\sigma_b'=-\sigma_0 ~g \int\limits_0^\pi \frac{d\theta}{\pi}
\int\limits_0^\infty \frac{dx\cos\theta
(x+\cos\theta)}{(1-x^2)\sqrt{x^2+1-2x\cos\theta}},
\end{eqnarray}
where the integral is taken in the principal value sense. Using the
identity
\begin{equation}
\int\limits_0^\infty
\frac{dx(x+\cos\theta)}{(1-x^2)\sqrt{x^2+1-2x\cos\theta}}=\frac{\sin^2(\theta/2)}{\cos{(\theta/2)}}\ln{[\tan({\theta}/{4})]},
\end{equation}
and integrating over the angle $\theta$ we obtain,
\begin{equation}
\label{vertex} \sigma_b'=\sigma_0 ~g\frac{8-3\pi}{6}.
\end{equation}
Combining Eqs.~(\ref{self}) and (\ref{vertex}) we finally arrive at
Eq.~(\ref{result}). The vertex correction is negative and nearly
cancels the self energy correction. The frequency independence of
Eqs.~(\ref{self}) and (\ref{vertex}) is analogous to the
independence of the non-interacting conductivity $\sigma_0$.

\section{Kubo formula for conductivity} An advantage  of
deriving minimal conductivity from the polarization operator originates from the fact that large logarithmic contributions {\it do not
appear} in different terms in this formalism. On the other hand one
could begin with a straightforward application of the Kubo formula,
which as we will see, does not offer such a simplification. As a result
one has to deal with logarithmic contributions which ultimately
cancel. The starting expression is the expression for the optical
conductivity
\begin{equation}
\label{kubo} \widetilde
\sigma(\omega)=\frac{K(\omega)-K(0)}{\omega},
\end{equation}
via the current-current correlation function, which is given in the
zeroth order by
\begin{equation}
\label{k0} K_0(\omega)= 4e^2v^2 \mbox{Tr} \sum_{\bf p}
\int\frac{d\epsilon }{2\pi} ~ \sigma_x \hat G_{\epsilon \bf p} \hat
\sigma_x \hat G_{\epsilon+\omega, \bf p},
\end{equation}
and in the first order by
\begin{eqnarray}
\label{ka} K_a(\omega)&=& 8ie^2v^2 \mbox{Tr} \sum_{\bf p p'}
\int\frac{d\epsilon d\epsilon'}{(2\pi)^2} ~V_{\bf p-p'} \nonumber\\
&& \times \hat \sigma_x \hat G_{\epsilon \bf p}
\hat G_{\epsilon' \bf p'}\hat \sigma_x \hat  G_{\epsilon \bf p} \hat G_{\epsilon+\omega, \bf p},\\
\label{kb} K_b(\omega)&=& 4ie^2v^2 \mbox{Tr} \sum_{\bf p p'}
\int\frac{d\epsilon d\epsilon'}{(2\pi)^2} ~V_{\bf p-p'} \nonumber\\
&& \times  \hat \sigma_x \hat  G_{\epsilon \bf p} \hat G_{\epsilon'
\bf p'}\hat \sigma_x \hat  G_{\epsilon'+\omega, \bf p'} \hat
G_{\epsilon+\omega, \bf p},
\end{eqnarray}
here the subscripts ${a,b}$ denote the contributions  from the self
energy and vertex diagrams, Fig.~1. Note, however, that the
corresponding contributions into the conductivity, which we denote
here by $\widetilde \sigma_a$ and $\widetilde \sigma_b$, do not
satisfy the condition (\ref{conduc}) {\it term by term}.  However
their {\it sum has to obey it}, $\widetilde \sigma_a+\widetilde
\sigma_b=\sigma_a+\sigma_b$.

Calculation of Eqs.~(\ref{k0}-\ref{kb}) is similar to  the above
derivation for the polarization operator.
\begin{equation}
\label{kubo_a} \widetilde \sigma_{a}(\omega)=ie^2\omega \sum_{\bf p
p'}V_{\bf p-p'} \cos{\theta} \frac{12v^2p^2 -\omega^2}{p^2
(\omega^2-4v^2p^2)^2}, \end{equation}
\begin{equation}
 \label{kubo_b} \widetilde
\sigma_{b}(\omega)=ie^2 \omega \sum_{\bf p p'} V_{\bf
p-p'}\cos\theta
\frac{\frac{1}{pp'}(\omega^2-8v^2p^2)\cos\theta-4v^2}{(\omega^2-4v^2p^2)(\omega^2-4v^2p'^2)},
\end{equation}
where we omit the subscripts in $\cos\theta_{\bf p,p'}$. The
expressions (\ref{kubo_a},\ref{kubo_b}) are to be contrasted with
Eqs.~(\ref{a}) and (\ref{b}). The obvious distinction arises from
the fact that the integrals in Eqs.~(\ref{kubo_a}) and
(\ref{kubo_b}) are logarithmically divergent though these
divergencies ultimately cancel in their sum, $\widetilde \sigma
(\omega)=\widetilde \sigma_a (\omega)+\widetilde \sigma_b (\omega)$.
However, a more striking observation can be made if one calculates
the difference of the two expressions,
\begin{eqnarray}
\sigma(\omega) -\widetilde \sigma (\omega) =I,
\end{eqnarray}
where
\begin{eqnarray}
\label{I} I=2ie^2\omega\sum_{\bf p p'}V_{\bf
p-p'}\cos{\theta}\frac{p'-p\cos{\theta}}{p^2p'(\omega^2-4v^2p^2)}.
\end{eqnarray}
Before addressing the issue of a numerical value of $I$ let us
briefly describe the third method for the calculation of the
conductivity.

\section{Kinetic equation} The kinetic equation to the lowest order
in electron-electron interaction and in the presence of electric
filed has the form \cite{2}
\begin{equation}
\label{kinur} \frac{\partial \hat f_{\bf p}}{\partial t}+ivp [\hat{
\sigma}_{\bf p},\hat f_{\bf p}] +e{\bf E}\cdot\frac{\partial \hat
f_{\bf p}}{\partial \bf p}=i \sum_{\bf p'} V_{\bf p-p'} [\hat f_{\bf
p'},\hat f_{\bf p}],
\end{equation}
where $\hat f_{\bf p}$ is the $2\times 2$ matrix distribution
function.  {The second term in the left-hand side represents the
rate of change of the electron distribution function during its
precession in the momentum-dependent ``pseudo-Zeeman'' field. The
third term is the usual drift in the momentum space caused by
external electric field. Finally, the right-hand side account for
the exchange electron-electron interaction (Hartree contribution
being zero by virtue of electric neutrality and spatial
homogeneity).} Given the solution of kinetic equation (\ref{kinur})
one can find electric current and optical conductivity from ${\bf j}
=\sigma(\omega) {\bf E}= 4ev \mbox{Tr} \sum_{\bf p}  \hat f_{\bf p}
\hat {\bm \sigma} $. The detailed solution of Eq.~(\ref{kinur}) to
the first order in ${\bf E}$ and $V_{\bf p-p'}$ was found in
Ref.~\cite{2}. For the conductivity it yields,
\begin{eqnarray}
\label{kin} \sigma_{kin}(\omega)=8ie^2\omega \sum_{\bf p p'}V_{\bf
p-p'} \cos{\theta} \frac{v^2}{(\omega^2-4v^2p^2)^2}\nonumber\\-4i
e^2 \omega \sum_{\bf p p'} V_{\bf p-p'} \frac{v^2 p \cos^2\theta
+v^2 p' \cos\theta}{p'(\omega^2-4v^2p^2)(\omega^2-4v^2p'^2)}.
\end{eqnarray}
This expression is different from both $\sigma(\omega)$ and
$\widetilde \sigma(\omega)$. Interestingly,
\begin{eqnarray}
\label{kinI} \sigma(\omega) - \sigma_{kin} (\omega) =I/2.
\end{eqnarray}

\section{Discussion} Three different values obtained from the polarization operator, $\sigma(\omega)$, Kubo formula, $\widetilde
\sigma(\omega)$, and kinetic equation, $\sigma_{kin}(\omega)$,
respectively, indicate an inconsistency of the theory of
interacting two-dimensional Dirac fermions {\it unless}  $I=0$. We will
now demonstrate that the conclusion of whether $I=0$ or $I\ne 0$
depends on the way the ultraviolet cut-off is imposed in the
calculation of a singular integral over ${\bf p'}$ in Eq.~(\ref{I}).

(i) {\it Hard cut-off}. Let us first assume that the divergent
momentum integral is extended only to $p' \le {\cal K}$. By noting
that interaction potential depends only on $s=({\bf p}-{\bf p'})^2$
and that $p'-p\cos\theta =\frac{1}{2} \partial s/\partial p'$ we
then obtain for the latter integral in case when $V(s)=2\pi
e^2/\sqrt{s}$,
\begin{eqnarray}
\sum_{\bf p'}V_{\bf
p-p'}\cos{\theta}\frac{p-p\cos{\theta}}{p'}=
\frac{e^2}{2}\int\limits_0^{2\pi}\frac{d\theta}{2\pi}\cos{\theta}\int\limits_0^{\cal K}
 \frac{d p'}{\sqrt{s}} \frac{\partial s}{\partial p'}\nonumber\\ =
 e^2\int\limits_0^{2\pi}\frac{d\theta}{2\pi}\cos{\theta}
\sqrt{s({\cal K},p)}.
\end{eqnarray}
Expanding $\sqrt{s({\cal K},p)}\approx {\cal K}-p\cos\theta$ for
large values of ${\cal K}$, we observe that the integral here is
cut-off independent and equals $-e^2p/2$. It is now straightforward
to verify that equation (\ref{I}) gives $I=-\sigma_0g/2$. Such value
of $I$ yields Eq.~(\ref{Cher}) reproducing the result of
Ref.~\cite{herbut}, and precisely accounts for the difference
between Eq.~(\ref{Cher}) and our result (\ref{result}). However, as
shown above, such an ultraviolet cut-off yields {\it three
different} values of the conductivity depending on which method is
being used and is therefore {\it unphysical}.

(ii) {\it Soft cut-off}. The anomaly encountered in the  expression
(\ref{I}) is specific for $V_q \propto q^{-1}$ behavior of the
interaction potential. For {\it any} faster decay of interaction at
large momenta the integral $I$ {\it vanishes}.  Let us demonstrate
this point by assuming
\begin{equation}
\label{soft} V(q)=\frac{2\pi e^2}{q}e^{- q/{\cal K}},~~~ {\cal K}
\to \infty.
\end{equation}
Calculation similar to the preceding one gives,
\begin{equation}
\lim_{{\cal K} \to
\infty}\frac{e^2}{2}\int\limits_0^{2\pi}\frac{d\theta}{2\pi}\cos{\theta}\int\limits_0^\infty
d p' \frac{e^{- \sqrt{s}/{\cal K}}}{\sqrt{s}} \frac{\partial
s}{\partial p'}=0,
\end{equation}
so that $I=0$ and all three methods yield {\it the same} value
(\ref{result}). Similar conclusion will be reached if one assumes
$V(q)\propto q^{-1-\eta}$ and subsequently takes the limit $\eta \to
0$.

 { Having established that the hard cut-off utilized in
Ref.~\cite{herbut} in the course of Kubo calculations actually
results in different and hence inconsistent results when other
methods are used, it is time now to discuss the origin of this
inconsistency. Terminating momentum integrals at some value $p={\cal
K}$
 means in fact an essential modification of electron spectrum at large momenta
that effectively excludes these states from possible virtual
processes. Such a procedure, though not necessarily incorrect, can
be made self-consistent {\it only} if it is accompanied by the
appropriate change in the operators of electric current. Otherwise,
the Ward identity, which ensures particle conservation, is violated.
This is why the polarization function method, which does not involve
current vertices in the course of calculations, gives results (in
the form of convergent integrals) {\it independent} of the cut-off
procedure. On the other hand both the Kubo formalism and kinetic
equation do involve current operators explicitly and thus {\it fail}
if the hard cut-off is implemented without a proper modification of
current vertices.

To the contrary, the soft cut-off procedure presented in this Letter
does not require modifications of the electron spectra (Green's
functions) nor of the electric current vertices. It is thus
self-consistent and quite naturally yields {\it identical} values
for the conductivity irrespective of the method used.

}

\section{Conclusion}

We have calculated the first order interaction correction to the
conductivity of intrinsic graphene. Within the Kubo and kinetic
equation formalisms the self-energy and vertex corrections contain
large logarithmic frequency-dependent terms which ultimately cancel
each other. Within the more convenient approach based on the
calculation of the polarization operator, such terms do not appear
at all. Such a simplification originates from a simpler scalar
vertex in the case of a polarization operator.

Nevertheless, the three methods discussed in the present Letter
result in different, and hence, unphysical values for the
interaction correction unless the large-momentum cut-off is imposed
in the form of Eq.~(\ref{soft}), or similar. In that case all
methods yield the same value given by Eq.~(\ref{result}).

To summarize, the calculations presented above indicate that the
effects of electron-electron interactions  lead to finite though
numerically very small corrections to the minimal conductivity.
 { Finally, the calculations of the present Letter are
performed in the limit of zero temperature and their validity
implies that $\hbar \omega \gg k_BT$.}

\acknowledgments Many useful discussions with S.~Gangadharaiah, D.
Maslov, M. Raikh, A. Shytov, J. Schmalian, P. Silvestrov and O.
Starykh  are gratefully acknowledged. This work was supported by
DOE, Office of Basic Energy Sciences,
 Award No.~DE-FG02-06ER46313.

\newpage  \centerline{\bf Appendix}
\bigskip

After publication of our paper in Europhys.~Lett.~{\bf 83}, 17005
(2008), a preprint by I.F. Herbut, V. Juricic, O. Vafek, and M.J.
Case, {\it \bf "Comment on "Minimal conductivity in graphene:
Interaction corrections and ultraviolet anomaly" by Mishchenko E.
G."}, appeared in arXiv:0809.0725. It was previously under
consideration for publication in the Europhysics Letters.
Below we present our reply:\\

The Comment argues against the procedure implemented above, which
leads to Eq.~(\ref{result}), and advocates dimensional
regularization scheme in support of the value, Eq.~(\ref{Cher}),
obtained in Ref.~\cite{herbut}. Yet, the Comment fails to offer a
consistent resolution of the issue. Indeed, following our suggestion
to utilize the charge conservation law, Eq.~(\ref{conduc}), the
authors of the Comment analyzed the derivation of the interaction
corrections from the polarization operator and reported that the
dimensional regularization yielded, $C=(11 - 3\pi)/6\approx 0.26$,
the value {\it different} from their Eq.~(\ref{Cher}). (Note that
this value coincides with $\sigma_{kin}$ given by Eq.~(\ref{kinI})
of the present paper when $I$ is calculated with the help of the
hard cut-off.) Addressing this discrepancy, the authors of the
Comment conclude only that, "The origin of this non-uniqueness is
unclear at the moment, but we suspect that it may be the non-gauge
invariant contribution to the conductivity which is not properly
treated within the density polarization approach."

Citing some unidentified contribution does not add clarity to the
discussion. The equivalence between the Kubo and the density
polarization approaches in the calculation of the homogeneous
conductivity is {\it ensured by the charge conservation law} (the
Ward identity). It is surprising that the authors end the discussion
with the above statement and do not even attempt to find out what
happens to the charge conservation in their calculations. It is thus
difficult to conclude that the authors of the Comment were able "to
clarify and correct some of the statements made" in our paper. If
anything, the credibility of their result, Eq.~(\ref{Cher}), is even
more questionable as it is now clear that this result is based on
the scheme that yields values which vary depending on the method
used.

Interestingly, recent measurements of dynamic conductivity
\cite{Nair} show $\sigma=\sigma_0 (1.01\pm 0.04)$ over visible
frequencies range and thus point towards smaller values of
interaction corrections.


\begin{thebibliography}{50}


\bibitem{fradkin} E. Fradkin, Phys. Rev. B {\bf 33}, 3263 (1986).

\bibitem{lee}
P. A. Lee, Phys. Rev. Lett. {\bf 71}, 1887 (1993).
\bibitem{ludwig}
A. W. W. Ludwig, M. P. A. Fisher, R. Shankar, and G. Grinstein,
Phys. Rev. B {\bf 50}, 7526 (1994).

\bibitem{shon} N. H. Shon and T. Ando, J. Phys. Soc. Jpn. {\bf 67}, 2421 (1998).

\bibitem{gusynin} V. P. Gusynin and S. G. Sharapov, Phys. Rev. Lett. {\bf 95}, 146801
(2005).


\bibitem{kats} M. I. Katsnelson, Eur. J. Phys. B {\bf 51}, 157 (2006).

\bibitem{jakub} J. Tworzydlo, B. Trauzettel, M. Titov, A. Rycerz, C.W.J.
Beenakker, Phys. Rev. Lett. {\bf 96}, 246802 (2006).

\bibitem{cserti} J.~Cserti, Phys. Rev. B {\bf 75}, 033405 (2007).



\bibitem{nomura} K. Nomura and A. H. MacDonald, Phys. Rev. Lett. {\bf 98}, 076602 (2007).

\bibitem{falk} L. A. Falkovsky and A. A. Varlamov, Eur. Phys. J. B {\bf 56}, 281 (2007).

\bibitem{peres} N. M. R. Peres, F. Guinea, and A. H. Castro Neto, Phys. Rev. B
{\bf 73}, 125411 (2006).

\bibitem{ziegler}
K. Ziegler, Phys. Rev. Lett. {\bf 97}, 266802 (2006).

\bibitem{akira} S. Ryu, C. Mudry, A. Furusaki, and A.W.W. Ludwig, Phys. Rev. B {\bf 75}, 205344
(2007).

\bibitem{Geim} K. S. Novoselov, A. K. Geim, S. V. Morozov, D. Jiang, Y.
Zhang, S. V. Dubonos, I. V. Grigorieva, and A. A. Firsov, Science
{\bf 306}, 666 (2004).

\bibitem{Nov} K. S. Novoselov, A. K. Geim, S. V. Morozov, D. Jiang, M. I.
Katsnelson, I. V. Grigorieva, S. V. Dubonos, and A. A. Firsov,
Nature {\bf 438}, 197 (2005); A. K. Geim and K. S. Novoselov, Nature
Materials {\bf 6}, 183 (2007).

\bibitem{2} E.G. Mishchenko, Phys. Rev. Lett. {\bf 98}, 216801 (2007).

\bibitem{sheehy} D.E. Sheehy and J. Schmalian, Phys. Rev. Lett. {\bf 99}, 226803 (2007).

\bibitem{herbut} I.F. Herbut, V. Juricic, and O. Vafek,
Phys. Rev. Lett. {\bf 100}, 046403 (2008).




\bibitem{note} The polarization operator in the non-interacting system is given by
$\Pi_0(\omega,q)=-4i\mbox{Tr} \int \frac{d\epsilon}{2\pi} \sum_{\bf
p} \hat G_{\epsilon \bf p}\hat G_{\epsilon+\omega, \bf p+q}$, which
yields
$$
\sigma_0=2ie^2\omega \sum_{\bf p}\frac{v} {p~(\omega^2-4v^2p^2)},
$$

\bibitem{Gonzales} J. Gonz\'alez, F. Guinea, and M.A.H. Vozmediano,
Nucl. Phys. B {\bf 73}, 125411 (1994).

\bibitem{Gonzales1} J. Gonz\'alez, F. Guinea, and M.A.H. Vozmediano,
Phys. Rev. B {\bf 59}, R2474 (1999).

\bibitem{Nair} R.R. Nair, P. Blake, A.N. Grigorenko, K.S. Novoselov, T.J. Booth, T. Stauber, N.M.R. Peres, and A.K.
Geim,  Science {\bf 320}, 1308 (2008).


\end{thebibliography}
\end{document}